\documentclass
[aps,prl,amsfonts,amssymb,twocolumn,amsmath,preprintnumbers,nofootinbib,floatfix,
showpacs]{revtex4-1}%
\usepackage[dvips]{graphics}
\usepackage{graphicx}
\usepackage{bm}
\usepackage{amsmath}
\usepackage{amsfonts}
\usepackage{amssymb}
\usepackage{xcolor}
\usepackage{subfigure}
\usepackage{hyperref,hypcap}
\usepackage{braket}
\usepackage{commath}%
\providecommand{\U}[1]{\protect\rule{.1in}{.1in}}

\begin{document}

\title{Conserved spin current for the Mott relation}
\author{Cong Xiao, Jihang Zhu, Bangguo Xiong, and Qian Niu}
\affiliation{Department of Physics, The University of Texas at Austin, Austin, Texas 78712, USA}

\begin{abstract}
The conserved bulk spin current [PRL \textbf{96}, 076604 (2006)] defined as
the time derivative of the spin displacement operator ensures automatically
the Onsager relation between the spin Hall effect (SHE)\ and the inverse SHE.
Here we reveal another desirable property of this conserved spin current: the
Mott relation exists linking the SHE and its thermal-counterpart -- spin
Nernst effect (SNE). According to the Mott relation, the SNE is known once the
SHE is understood. In the two-dimensional Dirac-Rashba system with smooth
scalar disorder-potential, we find a sign change of the spin Nernst
conductivity when tuning the chemical potential.
\end{abstract}
\maketitle

In the rapidly extending fields of spintronics and spin-caloritronics, the
spin Hall effect (SHE) \cite{Sinova2015,Nagaosa2008} and its
thermal-counterpart -- spin Nernst effect (SNE)
\cite{Ma2010,Dyrdal2016,Borge2013,SNE2014,Tauber2012,Meyer2017,Sheng2017,Kim2017,Bose2018}
have played important roles. When describing the SHE and SNE in terms of the
bulk spin current in the presence of band-structure spin-orbit interaction,
there exists the well-known ambiguity about the definition of a transport spin
current when the transported spin-component is not conserved. A conserved bulk
spin current has been proposed by Shi, Zhang, Xiao and Niu (hereafter we call
it the SZXN spin current) \cite{Shi2006,note-conserve} and been studied
intensively
\cite{Zhang2008,Wong2008,Guo2009,Chen2014,Freimuth2010,Sugimoto2006,Mandal2008,Gorini2012}%
. The SZXN spin current operator is described as the time derivative of the
spin displacement operator (see below). The so-defined spin current has a
natural conjugate force and represents a transport current. The Onsager
relation can thus be established automatically between the SHE and inverse SHE
of this SZXN spin current \cite{Shi2006,Zhang2008,Gorini2012}. In this Rapid
Communication we reveal another desirable property of the SZXN spin current:
the Mott relation between the SHE and SNE. The Mott relation \cite{Streda1977}
can be viewed as a fundamental link between the transport current responses to
the electric field and to the temperature gradient in independent-carrier
systems with elastic scattering off disorder. According to the Mott relation,
the SNE is known once the SHE is understood.

As applications, we show that, in the weak disorder-potential regime, both the
SHE and SNE can be finite in the two-dimensional (2D) Dirac-Rashba system with
smooth disorder-potential, contrary to the vanishing SHE
\cite{Sugimoto2006,Mandal2008} and SNE in a Rashba 2D electron gas. A sign
change of the spin Nernst conductivity is found when tuning the chemical potential.

\emph{Generalized Mott relation}--- The out-of-equilibrium average value of an
observable $\hat{O}$ in a single-particle system reads $\delta O=$
Tr$\left\langle \hat{O}^{eq}\left(  \delta\hat{\rho}\right)  \right\rangle +$
Tr$\left\langle \hat{\rho}^{eq}\delta\hat{O}\right\rangle $ in the linear
response regime. Here $\hat{\rho}$ is the single-particle density matrix with
$\hat{\rho}^{eq}$ and $\delta\hat{\rho}$ the equilibrium and linear-response
components, respectively. $\hat{O}^{eq}$ and $\delta\hat{O}$ have analogous
meanings, $\left\langle ..\right\rangle $ denotes the disorder averaging. The
usual external perturbations driving nonequilibrium steady-states in
experiments are electric field $\mathbf{E}$ and temperature gradient
$-\mathbf{\nabla}T/T$. For transport effects, the temperature gradient can be
equivalently replaced by the gradient $-\mathbf{\nabla}\psi/c^{2}$ of a
fictitious gravitational potential $\psi$\ introduced by Luttinger (c is the
speed of light) \cite{Luttinger1964}. The first term of $\delta O$ arises from
the density-matrix linear response $\delta\hat{\rho}=\delta^{\mathbf{E}}%
\hat{\rho}+\delta^{\psi}\hat{\rho}$, whereas the second term comes from the
linear response of the observable operator itself with respect to external
perturbations $\delta\hat{O}=\delta^{\mathbf{E}}\hat{O}+\delta^{\psi}\hat{O}$.
As a result, the linear response of any transport current $\hat{O}$ of a
single-electron system with respect to the d.c. uniform $\mathbf{E}$ and
$-\mathbf{\nabla}\psi/c^{2}$ reads ($\alpha,\beta=x,y$)%
\begin{equation}
\delta O_{\alpha}=L_{\alpha\beta}^{oe}E_{\beta}+L_{\alpha\beta}^{oQ}\left(
\frac{-\partial_{\beta}\psi}{c^{2}}\right)  , \label{linear response}%
\end{equation}
where $L_{\alpha\beta}^{o\xi}=D_{\alpha\beta}^{o\xi}+M_{\alpha\beta}^{o\xi}$
with $D_{\alpha\beta}^{oe}E_{\beta}\equiv$ Tr$\left\langle \hat{O}_{\alpha
}^{eq}\delta^{\mathbf{E}}\hat{\rho}\right\rangle $, $D_{\alpha\beta}%
^{oQ}\left(  \frac{-\partial_{\beta}\psi}{c^{2}}\right)  \equiv$
Tr$\left\langle \hat{O}_{\alpha}^{eq}\delta^{\psi}\hat{\rho}\right\rangle $,
$M_{\alpha\beta}^{oe}E_{\beta}\equiv$ Tr$\left\langle \hat{\rho}^{eq}%
\delta^{\mathbf{E}}\hat{O}_{\alpha}\right\rangle $ and $M_{\alpha\beta}%
^{oQ}\left(  \frac{-\partial_{\beta}\psi}{c^{2}}\right)  \equiv$
Tr$\left\langle \hat{\rho}^{eq}\delta^{\psi}\hat{O}_{\alpha}\right\rangle $.
The basic considerations for obtaining $\delta^{\mathbf{E,}\psi}\hat{\rho}$
and $\delta^{\mathbf{E,}\psi}\hat{O}$ can be found in the classical treatment
in Ref. \cite{Streda1977}, where the electric field enters the total
single-carrier Hamiltonian $\hat{H}^{t}$\ via the dipole term $-e\mathbf{\hat
{r}\cdot E}$. This is the case in the level of the full Hamiltonian
\cite{Chang2008,Xiao2018effective}. While in the level of an effective
Hamiltonian, the canonical position $\mathbf{\hat{r}}$ may not be the physical
one $\mathbf{\hat{r}}^{phy}$ and an anomalous dipole $e\left(  \mathbf{\hat
{r}}^{phy}-\mathbf{\hat{r}}\right)  $\ (usually related to effective
spin-orbit interaction \cite{Sinitsyn2008,Culcer2013}) couples to the electric
field \cite{Chang2008}. This situation needs separate treatment
\cite{Culcer2013,Xiao2018effective}. In the present study we neglect this
complexity and take $\mathbf{\hat{r}}^{phy}=\mathbf{\hat{r}}$ approximately
even when the transport is calculated in the level of effective Hamiltonians
\cite{note-appro}. Thus the spin-orbit interactions with the external electric
field and with the disorder potential
\cite{Chang2008,Xiao2018effective,Culcer2013} do not appear throughout this
Rapid Communication.

$D_{\alpha\beta}^{o\xi}$ is generally given by the Bastin version of $\hat
{O}_{\alpha}-\hat{\jmath}_{\beta}^{\xi}$ correlation function
\cite{Bastin1971}, which can be casted into \cite{Bruno2001}%
\begin{equation}
D_{\alpha\beta}^{o\xi}=D_{\alpha\beta}^{o\xi,I\left(  a\right)  }%
+D_{\alpha\beta}^{o\xi,I\left(  b\right)  }+D_{\alpha\beta}^{o\xi,II}%
\end{equation}
with
\[
D_{\alpha\beta}^{o\xi,I\left(  a\right)  }=-\frac{\hbar}{2\pi}\int
d\epsilon\frac{df^{0}\left(  \epsilon\right)  }{d\epsilon}\text{Tr}%
\left\langle \hat{O}_{\alpha}^{eq}\hat{G}^{R}\left(  \epsilon\right)
\hat{\jmath}_{\beta}^{eq,\xi}\hat{G}^{A}\left(  \epsilon\right)  \right\rangle
,
\]%
\[
D_{\alpha\beta}^{o\xi,I\left(  b\right)  }=\frac{\hbar}{2\pi}%
\mathrm{\operatorname{Re}}\int d\epsilon\frac{df^{0}\left(  \epsilon\right)
}{d\epsilon}\text{Tr}\left\langle \hat{O}_{\alpha}^{eq}\hat{G}^{R}\left(
\epsilon\right)  \hat{\jmath}_{\beta}^{eq,\xi}\hat{G}^{R}\left(
\epsilon\right)  \right\rangle ,
\]
and%
\begin{align*}
D_{\alpha\beta}^{o\xi,II}  &  =\frac{\hbar}{2\pi}\operatorname{Re}\int
d\epsilon f^{0}\left(  \epsilon\right)  \text{Tr}\left\langle \hat{O}_{\alpha
}^{eq}\hat{G}^{R}\left(  \epsilon\right)  \hat{\jmath}_{\beta}^{eq,\xi}%
\frac{d\hat{G}^{R}\left(  \epsilon\right)  }{d\epsilon}\right. \\
&  \left.  -\hat{O}_{\alpha}^{eq}\frac{d\hat{G}^{R}\left(  \epsilon\right)
}{d\epsilon}\hat{\jmath}_{\beta}^{eq,\xi}\hat{G}^{R}\left(  \epsilon\right)
\right\rangle .
\end{align*}
Here $\hat{\jmath}_{\beta}^{eq,\xi}$ stands for the equilibrium electric
current ($\xi=e$) and heat current ($\xi=Q$) operators: $\hat{\jmath}_{\beta
}^{eq,e}=e\hat{v}_{\beta}$, $\hat{\jmath}_{\beta}^{eq,Q}=\frac{1}{2}\left\{
\hat{H}^{eq}-\mu,\hat{v}_{\beta}\right\}  $. $\hat{G}^{R/A}\left(
\epsilon\right)  =\left(  \epsilon-\hat{H}^{eq}\pm i0^{+}\right)  ^{-1}$ with
$\hat{H}^{eq}$ the single-particle Hamiltonian at equilibrium, $f^{0}$ is the
equilibrium Fermi distribution and $\mu$ is the chemical potential.
\begin{widetext}
Now we derive a general relation between $D_{\alpha\beta}^{oQ}$ and
$D_{\alpha\beta}^{oe}$. By using $\left(  \hat{G}^{R/A}\right)  ^{2}=-d\hat
{G}^{R/A}/d\epsilon$, we get%
\[
D_{\alpha\beta}^{oQ,I\left(  a\right)  }=\int d\epsilon\left[  -\frac
{df^{0}\left(  \epsilon\right)  }{d\epsilon}\right]  \frac{\left(
\epsilon-\mu\right)  }{e}D_{\alpha\beta}^{oe,I\left(  a\right)  }\left(
T=0,\epsilon\right)  +\frac{\hbar}{4\pi}\int d\epsilon\frac{df^{0}\left(
\epsilon\right)  }{d\epsilon}\text{Tr}\left\langle \hat{O}_{\alpha}^{eq}%
\hat{v}_{\beta}^{eq}\hat{G}^{A}\left(  \epsilon\right)  +\hat{v}_{\beta}%
^{eq}\hat{O}_{\alpha}^{eq}\hat{G}^{R}\left(  \epsilon\right)  \right\rangle ,
\]%
\[
D_{\alpha\beta}^{oQ,I\left(  b\right)  }=\int d\epsilon\left[  -\frac
{df^{0}\left(  \epsilon\right)  }{d\epsilon}\right]  \frac{\left(
\epsilon-\mu\right)  }{e}D_{\alpha\beta}^{oe,I\left(  b\right)  }\left(
T=0,\epsilon\right)  -\frac{\hbar}{4\pi}\int d\epsilon\frac{df^{0}\left(
\epsilon\right)  }{d\epsilon}\text{Tr}\left\langle \frac{1}{2}\left\{  \hat
{O}_{\alpha}^{eq},\hat{v}_{\beta}^{eq}\right\}  \left[  \hat{G}^{R}\left(
\epsilon\right)  +\hat{G}^{A}\left(  \epsilon\right)  \right]  \right\rangle ,
\]
and%
\[
D_{\alpha\beta}^{oQ,II}=-\int d\epsilon\left[  f^{0}\left(  \epsilon\right)
+\left(  \epsilon-\mu\right)  \frac{df^{0}\left(  \epsilon\right)  }%
{d\epsilon}\right]  \frac{1}{e}D_{\alpha\beta}^{oe,II}\left(  T=0,\epsilon
\right)  +\frac{\hbar}{4\pi}\int d\epsilon\frac{df^{0}\left(  \epsilon\right)
}{d\epsilon}\text{Tr}\left\langle \frac{1}{2}\left[  \hat{O}_{\alpha}%
^{eq},\hat{v}_{\beta}^{eq}\right]  \left[  \hat{G}^{R}\left(  \epsilon\right)
-\hat{G}^{A}\left(  \epsilon\right)  \right]  \right\rangle ,
\]
then $D_{\alpha\beta}^{oQ}\left(  T,\mu\right)  =D_{\alpha\beta}^{oQ,I}\left(
T,\mu\right)  +D_{\alpha\beta}^{oQ,II}\left(  T,\mu\right)  $ yields the first
main result of this paper:
\begin{equation}
D_{\alpha\beta}^{oQ}\left(  T,\mu\right)  =\int d\epsilon\left[  -\frac
{df^{0}\left(  \epsilon\right)  }{d\epsilon}\right]  \frac{\left(
\epsilon-\mu\right)  }{e}D_{\alpha\beta}^{oe}\left(  T=0,\epsilon\right)
-\frac{1}{e}\int d\epsilon f^{0}\left(  \epsilon\right)  D_{\alpha\beta
}^{oe,II}\left(  T=0,\epsilon\right)  .\label{D}%
\end{equation}
On the other hand, utilizing $\left(  \hat{G}^{R/A}\right)  ^{2}=-d\hat
{G}^{R/A}/d\epsilon$ and \cite{Bruno2001} $i\hbar\hat{G}^{R}\hat{v}_{\beta
}^{eq}=\hat{G}^{R}\left[  \hat{r}_{\beta},\hat{H}^{eq}\right]  =\hat{G}%
^{R}\left[  \left(  \hat{G}^{R}\right)  ^{-1},\hat{r}_{\beta}\right]  $, we
get
\[
D_{\alpha\beta}^{oe,II}=\frac{e}{2}\int d\epsilon\frac{df^{0}\left(
\epsilon\right)  }{d\epsilon}\text{Tr}\left\langle \left\{  \hat{O}_{\alpha
}^{eq},\hat{r}_{\beta}\right\}  \delta\left(  \epsilon-\hat{H}^{eq}\right)
\right\rangle +\frac{e}{\pi}\operatorname{Im}\int d\epsilon f^{0}\left(
\epsilon\right)  \text{Tr}\left\langle \hat{O}_{\alpha}^{eq}\hat{G}^{R}\left(
\epsilon\right)  \hat{r}_{\beta}\hat{G}^{R}\left(  \epsilon\right)
\right\rangle .
\]
\end{widetext}We find that, if the current $\hat{O}_{\alpha}$ is defined in
terms of the time derivative of some displacement operators \cite{Shi2006},
i.e.,
\begin{equation}
\hat{O}_{\alpha}=\frac{1}{i\hbar}\left[  \hat{F}_{\alpha},\hat{H}^{t}\right]
\text{ where }\left[  \hat{F}_{\alpha},\hat{r}_{\beta}\right]  =0,
\label{condition}%
\end{equation}
then $Tr\left\langle \hat{O}_{\alpha}^{eq}\hat{G}^{R}\hat{r}_{\beta}\hat
{G}^{R}\right\rangle =\frac{1}{i\hbar}Tr\left\langle \left[  \hat{r}_{\beta
},\hat{F}_{\alpha}\right]  \hat{G}^{R}\right\rangle =0$ and%
\begin{equation}
D_{\alpha\beta}^{oe,II}=\frac{e}{2}\int d\epsilon\frac{df^{0}\left(
\epsilon\right)  }{d\epsilon}\text{Tr}\left\langle \left\{  \hat{O}_{\alpha
}^{eq},\hat{r}_{\beta}\right\}  \delta\left(  \epsilon-\hat{H}^{eq}\right)
\right\rangle . \label{Streda term}%
\end{equation}

For the current $\hat{O}_{\alpha}$ in the form of Eq. (\ref{condition}), one
has $\delta^{\mathbf{E}}\hat{O}=0$ and $\delta^{\psi}\hat{O}=\frac{1}%
{2}\left\{  \hat{r}_{\beta}\mathbf{,}\hat{O}^{eq}\right\}  \frac
{\partial_{\beta}\psi}{c^{2}}$, thus $\delta O_{\alpha}=D_{\alpha\beta}%
^{oe}E_{\beta}+\left(  D_{\alpha\beta}^{oQ}+M_{\alpha\beta}^{oQ}\right)
\left(  \frac{-\partial_{\beta}\psi}{c^{2}}\right)  $, where $M_{\alpha\beta
}^{oQ}\left(  \frac{-\partial_{\beta}\psi}{c^{2}}\right)  \equiv$
Tr$\left\langle \hat{\rho}^{eq}\delta^{\psi}\hat{O}_{\alpha}\right\rangle $ is
given by \cite{Streda1977}
\begin{align}
M_{\alpha\beta}^{oQ}  &  =-\frac{1}{2}\int d\epsilon f^{0}\left(
\epsilon\right)  \text{Tr}\left\langle \delta\left(  \epsilon-\hat{H}%
^{eq}\right)  \left\{  \hat{r}_{\beta},\hat{O}_{\alpha}^{eq}\right\}
\right\rangle \nonumber\\
&  =\frac{1}{e}\int d\epsilon f^{0}\left(  \epsilon\right)  D_{\alpha\beta
}^{oe,II}\left(  T=0,\epsilon\right)  . \label{magnetization}%
\end{align}
Combining Eqs. (\ref{Streda term}), (\ref{magnetization}) and (\ref{D}) yields
the generalized Mott relation%
\begin{equation}
L_{\alpha\beta}^{oQ}\left(  T,\mu\right)  =\int d\epsilon\left[  -\frac
{df^{0}\left(  \epsilon\right)  }{d\epsilon}\right]  \frac{\left(
\epsilon-\mu\right)  }{e}L_{\alpha\beta}^{oe}\left(  T=0,\epsilon\right)
\label{Mott}%
\end{equation}
for the current $\hat{O}_{\alpha}$ having the form of Eq. (\ref{condition}).
This relation is exactly the same as the well-known generalized Mott relation
\cite{Streda1977} between $L_{\alpha\beta}^{eQ}$ and $L_{\alpha\beta}^{ee}$.
Equations (\ref{D}) - (\ref{Mott}) are the main result of this Rapid
Communication. When the distances between the chemical potential and the band
edges are much larger than the thermal energy $k_{B}T$, the Sommerfeld
expansion is legitimate \cite{Xiao2016PRB}, yielding the standard Mott
relation%
\begin{equation}
L_{\alpha\beta}^{oQ}\left(  T,\mu\right)  /T=\frac{\pi^{2}k_{B}^{2}T}{3e}%
\frac{\partial L_{\alpha\beta}^{oe}\left(  T=0,\epsilon\right)  }%
{\partial\epsilon}|_{\epsilon=\mu}, \label{Mott-1}%
\end{equation}
which relates $L_{\alpha\beta}^{oQ}$ to the energy derivative of
$L_{\alpha\beta}^{oe}$ around the chemical potential.

Both the electric current operator $\mathbf{\hat{\jmath}}^{e}=e\frac{1}%
{i\hbar}\left[  \mathbf{\hat{r}},\hat{H}^{t}\right]  $ and the SZXN spin
current operator $\mathbf{\hat{\jmath}}^{s}=\frac{1}{i\hbar}\left[
\mathbf{\hat{r}}\hat{s}_{z},\hat{H}^{t}\right]  $ have the form of Eq.
(\ref{condition}). Thus the SNE of the SZXN current can be obtained once its
SHE is known.

\emph{Applications}---The intrinsic spin Hall conductivity $\sigma_{yx}%
^{s,in}$ of the SZXN current can be obtained by standard Kubo formula
\cite{Shi2006,Zhang2008,Freimuth2010}. Aside from the intrinsic contribution,
there exists disorder-induced contribution to the SHE
\cite{Sinova2015,Nagaosa2008,Sugimoto2006,Mandal2008}. Among the several
mechanisms of the extrinsic contribution, the one arising from the
band-off-diagonal elements of the out-of-equilibrium single-carrier density
matrix \cite{KL1957,Luttinger1958} has attracted much recent attention
\cite{Sinitsyn2008,Culcer2017,Xiao2017SHE,Xiao2018scaling}. Resorting to the
density-matrix transport theory in the weak disorder-potential regime
\cite{KL1957,Culcer2017,supp} with well-defined multiband structure
\cite{Xiao2017SOT}, this mechanism contributes a spin current in the form
\cite{off-diagonal}%
\begin{equation}
\mathbf{j}^{s,ex}=\sum_{l}g_{l}^{\left(  -2\right)  }\mathbf{j}_{l}^{s,ex}.
\label{SHC-sj}%
\end{equation}
Here $g_{l}^{\left(  -2\right)  }$ is just the conventional out-of-equilibrium
distribution function in the Boltzmann transport theory, in the order of
$\left\langle V^{2}\right\rangle ^{-1}$ with $V$ the disorder potential.
$l=\left(  \eta,\mathbf{k}\right)  $ where $\eta$ is the band index and
$\mathbf{k}$ is the momentum. In the case of scalar disorder potential
$\left[  \mathbf{\hat{r}}\hat{s}_{z},\hat{V}\left(  \mathbf{\hat{r}}\right)
\right]  =0$, we get \cite{supp}%
\begin{equation}
\mathbf{j}_{l}^{s,ex}=\sum_{l^{\prime}}\omega_{l^{\prime}l}^{\left(  2\right)
}\left(  \mathbf{A}_{l^{\prime}}^{s}-\mathbf{A}_{l}^{s}\right)  \label{sj}%
\end{equation}
when $s_{z}^{l}\equiv\langle u_{l}|\hat{s}_{z}|u_{l}\rangle=0$. The expression
of $\mathbf{j}_{l}^{s,ex}$ in the case of $s_{z}^{l}\neq0$ is given in the
Supplemental Material \cite{supp}.\ Here $\mathbf{A}_{l}^{s}\equiv i\langle
u_{l}|\hat{s}_{z}|\partial_{\mathbf{k}}u_{l}\rangle$, $\mathbf{|}%
l\rangle\equiv|\mathbf{k}\rangle|u_{l}\rangle$ is the eigenstate of the
disorder-free equilibrium Hamiltonian $\hat{H}_{0}^{eq}$\ with energy
$\epsilon_{l}$, and $\omega_{l^{\prime}l}^{\left(  2\right)  }=\frac{2\pi
}{\hbar}\left\langle \left\vert V_{ll^{\prime}}\right\vert ^{2}\right\rangle
\delta\left(  \epsilon_{l}-\epsilon_{l^{\prime}}\right)  $ is the
lowest-Born-order scattering rate. Since $s_{z}^{l}=0$, $\mathbf{A}_{l}^{s}$
is real and remains unchanged under a local U(1) gauge transformation
$|u_{l}\rangle\rightarrow e^{i\phi_{l}}|u_{l}\rangle$. The extrinsic
contribution Eq. (\ref{SHC-sj}) can be independent of both the disorder
potential and impurity density, and thus may cancel partly or totally the
intrinsic SHE.

In the weak disorder-potential regime other disorder-induced contributions to
the SHE \cite{Sinitsyn2008,Xiao2017SHE,Xiao2018scaling,note-extrinsic} vanish
in the presence of weak scalar scattering when the Berry-curvatures on the
Fermi surfaces are zero. This can be appreciated most easily in the limit of
smooth disorder-potential varying slowly on the scale of the lattice constant
\cite{note}. Thus the disorder-induced SHE is just given by Eq. (\ref{SHC-sj}%
). This is the case in 2D systems with Rashba spin-orbit interaction, which
are the focus in the following model analysis.

We first apply above results to the 2D Rashba model\ (both Rashba subbands
partially occupied, Fig. 1(a)) with smooth scalar-impurity potentials,
arriving at vanishing SHE (Supplementary materials \cite{supp}) consistent
with previous works \cite{Sugimoto2006,Mandal2008}. According to the
generalized Mott relation, the SNE of the SZXN current
vanishes.\begin{figure}[b]
\includegraphics[width=1\columnwidth]{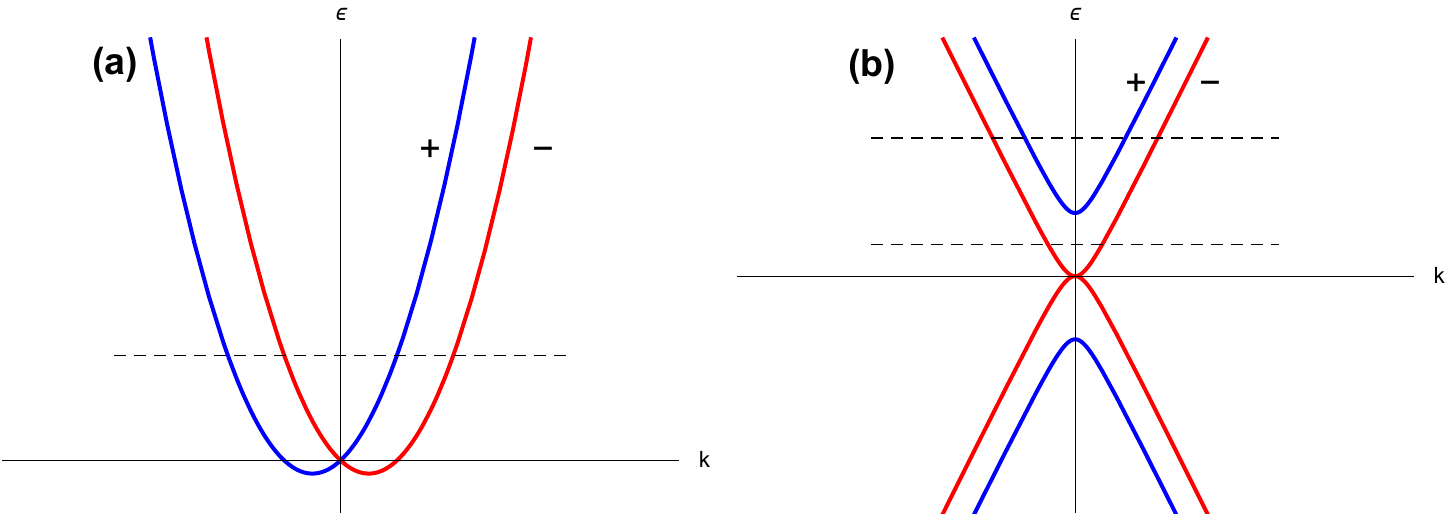}\caption{Schematic of the
band structures of the 2D Rashba model (a) and 2D Dirac-Rashba model (b).}%
\end{figure}

Now we discuss a model showing nonzero SHE and SNE of the SZXN current. As a
minimal model for low-energy electronic states around the Dirac point $K$ in a
graphene monolayer subject to $z\rightarrow-z$ asymmetric spin-orbit
interaction, the 2D Dirac-Rashba Hamiltonian in the A-B sublattice space reads
\cite{Rashba2009}%
\begin{align}
\hat{H}_{0}^{eq}  &  =v\left[
\begin{array}
[c]{cc}%
0 & \left(  k_{x}-ik_{y}\right)  \sigma_{0}\\
\left(  k_{x}+ik_{y}\right)  \sigma_{0} & 0
\end{array}
\right] \nonumber\\
&  +\lambda_{R}\left[
\begin{array}
[c]{cc}%
0 & \sigma_{y}+i\sigma_{x}\\
\sigma_{y}-i\sigma_{x} & 0
\end{array}
\right]  .
\end{align}
Here $v=\hbar v_{F}$, $\sigma_{i\text{ }}\left(  i=x,y,z\right)  $ is the
Pauli matrix and $\sigma_{0}$ the unit matrix in the spin space, $\lambda_{R}$
is the Rashba coupling. The four bands of $\hat{H}_{0}^{eq}$ read
$\epsilon_{k}^{\eta\zeta}=\eta\left(  \sqrt{\lambda_{R}^{2}+\left(  vk\right)
^{2}}+\zeta\lambda_{R}\right)  $. Here $\eta=\pm1$ denote conduction or
valence bands, $\zeta=\pm1$ denote spin subbands. We only consider the n-doped
case (Fig. 1(b)).

For the intrinsic SHE, a lengthy but straightforward calculation leads to the
results presented in Table I. In the presence of smooth scalar disorder
potential the intervalley scattering is suppressed, thus we obtain
\begin{equation}
\left(  \mathbf{j}_{l}^{s,ex}\right)  _{y}=-\frac{\hbar}{4}\frac{\hbar v_{F}%
}{\epsilon_{l}}\sin\xi\frac{1}{\tau_{l}^{tr}}\cos\phi
\end{equation}
in Eq. (\ref{SHC-sj}), where we use $\left(  \mathbf{A}_{\eta\zeta\mathbf{k}%
}^{s}\right)  _{y}=\eta\frac{\hbar}{4}\frac{\hbar v_{F}}{\epsilon_{l}}\sin
\xi\cos\phi$ ($s_{z}^{l}=0$ in this model) with $\sin\xi=vk/\sqrt{\lambda
_{R}^{2}+\left(  vk\right)  ^{2}}$ and $\tan\phi=k_{y}/k_{x}$. $\tau_{l}%
^{tr}=1/\sum_{l^{\prime}}\omega_{l^{\prime}l}^{\left(  2\right)  }\left[
1-\cos\left(  \phi^{\prime}-\phi\right)  \right]  $ is the transport time in
the case of smooth scalar disorder. The out-of-equilibrium distribution
function reads%
\begin{equation}
g_{l}^{\left(  -2\right)  }=\frac{e}{\hbar}\mathbf{E}\cdot\left(  -\partial
f^{0}/\partial\mathbf{k}\right)  \tau_{l}^{tr},
\end{equation}
thus Eq. (\ref{SHC-sj}) yields the extrinsic SHE listed in Table I. The total
spin Hall conductivity $\sigma_{yx}^{s}$ is positive-definite and depends on
the Fermi energy, as shown in Table I.\begin{table}[t]
\centering\label{table}
\begin{tabular}
[c]{|c|c|c|}\hline
& $\epsilon_{F}>2\lambda_{R}$ & $\epsilon_{F}<2\lambda_{R}$\\\hline
$\sigma_{yx}^{s,in}$ & $-\frac{3e}{8\pi}\frac{\epsilon_{F}^{2}-2\lambda
_{R}^{2}}{\epsilon_{F}^{2}-\lambda_{R}^{2}}$ & $-\frac{e}{16\pi}%
\frac{2\epsilon_{F}^{2}+\lambda_{R}\epsilon_{F}+2\lambda_{R}^{2}}{\lambda
_{R}\left(  \epsilon_{F}+\lambda_{R}\right)  }$\\\hline
$\sigma_{yx}^{s,ex}$ & $-\frac{e}{8\pi}\frac{\epsilon_{F}^{2}-2\lambda_{R}%
^{2}}{\epsilon_{F}^{2}-\lambda_{R}^{2}}$ & $-\frac{e}{16\pi}\frac{\epsilon
_{F}+2\lambda_{R}}{\epsilon_{F}+\lambda_{R}}$\\\hline
$\sigma_{yx}^{s}=\sigma_{yx}^{s,in}+\sigma_{yx}^{s,ex}$ & $-\frac{e}{2\pi
}\frac{\epsilon_{F}^{2}-2\lambda_{R}^{2}}{\epsilon_{F}^{2}-\lambda_{R}^{2}}$ &
$-\frac{e}{8\pi}\frac{\epsilon_{F}^{2}+\lambda_{R}\epsilon_{F}+2\lambda
_{R}^{2}}{\lambda_{R}\left(  \epsilon_{F}+\lambda_{R}\right)  }$\\\hline
\end{tabular}
\newline\caption{The intrinsic ($\sigma_{yx}^{s,in}$) and extrinsic
($\sigma_{yx}^{s,ex}$) spin Hall conductivities in the case of both conduction
bands partially occupied ($\epsilon_{F}>2\lambda_{R}$) and of empty inner
conduction band ($\epsilon_{F}<2\lambda_{R}$) in the 2D Dirac-Rashba model.}%
\end{table}

The spin Nernst conductivity $\alpha_{yx}^{s}=L_{yx}^{sQ}\left(  T,\mu\right)
/T$ is obtained by the Mott relations (\ref{Mott}) and (\ref{Mott-1}). In
particular, in the case of strong Rashba spin-orbit interaction, the chemical
potential may be located in the region $2\lambda_{R}\gg\mu\gg k_{B}T$
($\epsilon_{k=0}^{++}=2\lambda_{R}$) at low temperatures, then the standard
Mott relation (\ref{Mott-1}) applies, yielding
\begin{equation}
\frac{\alpha_{yx}^{s}}{T}=-\frac{\pi k_{B}^{2}}{24\lambda_{R}}\left[
1-\frac{3\lambda_{R}^{2}}{\left(  \mu+\lambda_{R}\right)  ^{2}}\right]  .
\end{equation}
This spin Nernst conductivity displays a sign change at $\mu/\lambda_{R}%
=\sqrt{3}-1$.

\emph{Discussion}---The SZXN spin current has been proved to obey the basic
near-equilibrium transport relations, i.e., the Mott relation established
above and the Onsager relation shown previously \cite{Shi2006,Zhang2008}. On
the other hand, for the conventional spin current defined as the
anti-commutator of the velocity and spin operators, whether the Mott relation
is valid or not (when the transported spin is non-conserved) is still a
problem not completely settled in literatures. Here we make some discussions
on this issue, because the conventional spin current is frequently used in
theoretical formulations of spin transport \cite{Sinova2015}, although it is
not directly related to the transport of spin in the case of spin
non-conservation \cite{Halperin2004}. Accordingly, in this case it is expected
that the Mott relation as a transport relation does not apply for the
conventional spin current. We point out that existing theories indeed do not
prove the Mott relation for the conventional spin current. Moreover, a recent
work showed the breakdown of the Mott relation for the conventional spin
current in a specific model \cite{Dyrdal2016}.

The direct application of the Kubo-Luttinger-Streda formalism presented in
this study to the thermoelectric response of the conventional spin current
does not yield the generalized Mott relation when the transported spin
component is not conserved. For the SNE of the conventional spin current, the
conventional-spin-current-heat-current correlation function reads \cite{supp}
($D_{yx}^{s0e}\equiv\sigma_{yx}^{s0}$)%
\begin{align}
D_{yx}^{s0Q}\left(  T,\mu\right)   &  =\int d\epsilon\left[  -\frac
{df^{0}\left(  \epsilon\right)  }{d\epsilon}\right]  \frac{\left(
\epsilon-\mu\right)  }{e}\sigma_{yx}^{s0}\left(  T=0,\epsilon\right)
\nonumber\\
&  -\frac{1}{e}\int d\epsilon f^{0}\left(  \epsilon\right)  \sigma
_{yx}^{s0,II}\left(  T=0,\epsilon\right)  .
\end{align}
However, $M_{yx}^{s0Q}\left(  T,\mu\right)  +D_{yx}^{s0Q}\left(  T,\mu\right)
$ cannot yield the Mott relation generally because $M_{yx}^{s0Q}\left(
T,\mu\right)  $ cannot be expressed as a Fermi sea integral of the so-called
\textquotedblleft Fermi sea term\textquotedblright%
\ \cite{Sinova2015,note-Fermisea} $\sigma_{yx}^{s0,II}\left(  T=0,\epsilon
\right)  $ of the conventional spin Hall conductivity. If one calculated only
the spin-current-heat-current correlation function $D_{yx}^{s0Q}$ and
neglected concurrently\ the Fermi sea term $\sigma_{yx}^{s0,II}$ of the spin
Hall conductivity, it would be concluded that the Mott relation is valid for
the conventional spin current. But this is not correct because generally both
of these two contributions are important \cite{Grimaldi2006,Streda1977}.

In the 2D Rashba model with scalar disorder, $\sigma_{yx}^{s0}=0$ \cite{Dimi}
and thus
\begin{equation}
D_{yx}^{s0Q}=-\frac{1}{e}\int d\epsilon f^{0}\left(  \epsilon\right)
\sigma_{yx}^{s0,II}\left(  T=0,\epsilon\right)  .
\end{equation}
The disorder-free part (dominates $\sigma_{yx}^{s0,II}$ in the weak
disorder-potential regime \cite{Sinova2015}) of\ $\sigma_{yx}^{s0,II}$ is
calculated to be $\sigma_{yx}^{s0,II}\left(  T=0,\epsilon\right)  =\frac
{e}{8\pi}\left(  \frac{k_{R}}{k_{0}\left(  \epsilon\right)  }-\frac
{k_{0}\left(  \epsilon\right)  }{k_{R}}\right)  \theta\left(  -\epsilon
\right)  $, with $\theta$ the step function and $k_{0}\left(  \epsilon\right)
=\alpha_{R}^{-1}\sqrt{\epsilon_{R}^{2}+2\epsilon_{R}\epsilon}$. Therefore, in
the low-temperature limit $D_{yx}^{s0Q}\left(  T\rightarrow0\right)
=-\frac{\epsilon_{R}}{12\pi T}$ is divergent when both Rashba subbands are
partially occupied. Recently, Dyrdal et al. \cite{Dyrdal2016} directly
evaluated the bubble \cite{Ma2010} and vertex corrections of $D_{yx}%
^{s0Q}\left(  T,\mu\right)  $ in the Rashba model, and obtained the same
low-temperature-limit value. They introduced a spin-resolved orbital
magnetization by hand and argued that this quantity also contributes a spin
current that should be added to the result of the
conventional-spin-current-heat-current correlation function \cite{Dyrdal2016}.
This treatment removes the divergent value of $D_{yx}^{s0Q}$ in the
zero-temperature limit in the Rashba model \cite{Dyrdal2016}, but yields a SNE
which does not follow the generalized Mott relation for the conventional spin current.

In summary, we proved the Mott relation for the spin thermoelectric transport
with the SZXN definition of the spin current. First-principle calculations of
the intrinsic SHE in terms of the SZXN current has been available in specific
materials such as some nonmagnetic hcp metals where the spin-nonconserving
part of the spin-orbit interaction could be important \cite{Freimuth2010}.
Thus the first-principle prediction of the intrinsic SNE according to the Mott
relation in these materials can be made.

\begin{acknowledgments}
We acknowledge insightful discussions with D. Li, Z. Ma, P. Streda, R. Raimondi, J. Borge and C. Gorini. C. X., B. X. and Q. N. are supported by DOE (DE-FG03-02ER45958, Division of Materials Science and Engineering), NSF (EFMA-1641101) and Welch Foundation (F-1255).
J. Z. is supported by the Welch foundation under Grant No. TBF1473.
\end{acknowledgments}


\begin{thebibliography}{99}
\bibitem {note-conserve}The SZXN spin current is conserved in the sense that
the field-generated spin torque does not appear in the spin continuity
equation, and the latter still includes a spin relaxation term.
\end{thebibliography}
\end{document}